\documentclass[dvipdfmx]{jpsj3}
\usepackage{txfonts}
\usepackage{bm}
\usepackage{amsmath,amssymb}
\usepackage{float}
\usepackage{ascmac}
\usepackage{subfigure}
\usepackage{verbatim}
\usepackage{wrapfig}
\usepackage{makeidx}
\usepackage{layout}
\usepackage{braket}
\usepackage{enumerate,amsmath,color,bm,ulem}
\newcommand{\rmd}{{\rm d}}

\newcommand{\iu}{{\rm i}}

\title{Nature of Driving Force on an Isolated Moving Vortex in Dirty Superconductors}
\author{Yusuke Kato$^{1}$\thanks{E-mail: yusuke@phys.c.u-tokyo.ac.jp} and Chun-Kit Chung$^{2}$\thanks{E-mail: ckchung@vortex.c.u-tokyo.ac.jp}}
\inst{
$^1$Department of Basic Science, The University of Tokyo, Tokyo 153-8902, Japan\\
$^2$Department of Physics, The University of Tokyo, Tokyo 113-0033, Japan}
\abst{We reconsider the force\textendash balance relation on an isolated vortex in the flux flow state within the scheme of the time\textendash dependent Ginzburg\textendash Landau (TDGL) equation. We define the force on a vortex by the total force on superconducting electrons in the region $S$ surrounding the vortex. We derive the local momentum balance relation of superconducting electrons and then find the force\textendash balance relation on an isolated vortex while taking account of the fact that the transport current in charged superconductors is inherently spatially varying with the scale of the penetration depth $\lambda$. We also find that the nature of the driving force is hydrodynamic when $S$ is a disk with radius $R$ satisfying $\xi\ll R\ll \lambda$ ($\xi$ is the coherence length) while the hydrodynamic and magnetic parts contribute equally to the driving force for $\lambda \lesssim R$.} 
\begin{document}
\maketitle
\noindent
{\it Introduction}: 
Vortex states in type II superconductors exhibit the resistive steady flow of quantum vortices in the presence of current density that exceeds a critical value\cite{KimParks,Tinkhamtext,Blatter}. The steady flow of a single vortex is often described by the force\textendash balance relation\cite{kopnintext}
\begin{equation}
\bm{j}_{\rm tr}\times \bm{\phi}^*_0 +\bm{F}_{\rm env}=0
\label{eq: f-b-relation-0}
\end{equation}
when the magnetic field is near the lower critical field $H_{\rm c1}$ and the intervortex spacing is larger than the penetration depth $\lambda$. 
Here, $\bm{j}_{\rm tr}$ denotes the transport current flowing through the superconductor and $\bm{\phi}^*_0$ is the vector with direction parallel to the magnetic field and modulus $|\phi^*_0|$ ($\phi^*_0=2\pi\hbar/e^*$ with $\hbar$ and $e^*$ being, respectively, the Planck constant divided by $2\pi$ and the electric charge of a Cooper pair). The first term in Eq.~(\ref{eq: f-b-relation-0}), $\bm{j}_{\rm tr}\times \bm{\phi}^*_0$, represents the force that drives the motion of the vortex and this force is to be balanced with $\bm{F}_{\rm env}$, the force due to impurities fixed to lattices or electron\textendash phonon interactions. 
This force $\bm{F}_{\rm env}$ is referred to as the environmental force in the literature (e.g., Ref.~\citen{kopnintext}) and  it is linear to the vortex velocity $\bm{v}_{\rm L}$ with accuracy $\mathcal{O}(\bm{v}_{\rm L}^2)$. 
The linear relation between $\bm{F}_{\rm env}$ and $\bm{v}_{\rm L}$, i.e., the transport coefficient, has been calculated within the scheme of the time\textendash dependent Ginzburg\textendash Landau (TDGL) theory\cite{schmid,GE68} for dirty superconductors\cite{GK71,huthompson,GK73a,GK73b,GK75,dorsey,kopnin93} (where the mean free path $l$ is much smaller than the coherence length $\xi$) or the equation of motion of the Gor'kov\textendash Keldysh Green function for clean superconductors ($l\gg \xi$ )\cite{kopnintext,LO86}. 
In these studies, the force $\bm{j}_{\rm tr}\times \bm{\phi}^*_0$ in Eq.~(\ref{eq: f-b-relation-0}) has often been called the Lorentz force. The nature of the force $\bm{j}_{\rm tr}\times \bm{\phi}^*_0$ is, however, not so obvious and there have been many discussions on this issue in the last fifty years\cite{KimParks,BS,NV,AoThouless,Sonin97}. 
At an early stage, Nozi\`eres and Vinen showed that a vortex flow is driven mainly by the Magnus force and not by the electromagnetic Lorentz force when $\lambda\gg \xi$\cite{NV}. The Magnus force on a superconducting vortex was intensively discussed in the 1990s\cite{AoThouless,Sonin97,Blatter}. The nature of the driving force on a vortex has, however, remained as an unsettled issue. 
With this background, what is important is to clarify the nature of the driving force within {\it an established scheme for the calculation of the transport coefficients of vortex dynamics}. We address this issue using the TDGL equation\cite{schmid,GE68} for dirty superconductors as a relatively simple case. Besides the nature of the driving force, we particularly emphasize that the following two points are important.
As far as we know, earlier studies on an isolated vortex in the flux flow state have assumed that the transport current is spatially uniform as in the case of neutral superfluids\cite{Sonin87}. Far away from a moving vortex, the current density $\bm{j}(\bm{r})$ should satisfy the London and Amp\`ere equations for the bulk superconductors and have the asymptotic form of 
\begin{equation}
\bm{j}(\bm{r})\sim \bm{j}_{\rm tr}(\bm{r})\equiv(j_+ {\rm e}^{y/\lambda}+j_- {\rm e}^{-y/\lambda})\bm{e}_x\label{eq:jtr-r}
\end{equation}
with the constants $j_\pm$ (we take the direction of the transport current as the $x$-axis).  
Once we take account of the spatial variation of the transport current, then the meaning of $\bm{j}_{\rm tr}$ in Eq.~(\ref{eq: f-b-relation-0}) should be stated more clearly.  One possibility is to read $\bm{j}_{\rm tr}$ in Eq.~(\ref{eq: f-b-relation-0}) as the spatially averaged value of the current $\langle\bm{j}(\bm{r})\rangle$. Another possibility is to replace $\bm{j}_{\rm tr}$ by the local value of the current density $\bm{j}(\bm{r}_{\rm L})$ or the supercurrent density $\bm{j}_{\rm s}(\bm{r}_{\rm L})$ at the position of the vortex center $\bm{r}_{\rm L}$.
Recall that the force on an isolated pinned vortex in the presence of supercurrent has the form of\cite{friedel,deGennes}
\begin{equation}
\bm{j}(\bm{r}_{\rm L})\times\bm{\phi}^*_0\label{eq: Jlocalphi}
\end{equation}
and has been reconsidered in Refs.~\citen{dxchen98} and \citen{Narayan}. 
In these references\cite{dxchen98,Narayan}, the authors concluded that Eq.~(\ref{eq: Jlocalphi}) is not the Lorentz force on a vortex on the basis of the London equation.  
As will be discussed in the rest of this paper, in the flux flow state of an isolated vortex, we find that neither Eq.~(1) nor (3) is correct but the force\textendash balance relation should be read as
\begin{equation}
\bm{j}_{\rm tr}(\bm{r}_{\rm L})\times \bm{\phi}^*_0 +\bm{F}_{\rm env}=0.
\label{eq: f-b-relation-4}
\end{equation}
$\bm{j}_{\rm tr}(\bm{r}_{\rm L})$ seems to be a local value but it is, in fact, an extrapolated value of $\bm{j}_{\rm tr}(\bm{r})$ at the vortex center. 

The other important point is the definition of the force; we define the force on an isolated vortex by the total force on the superconducting electrons in the region surrounding the vortex. This is how Nozi\`eres and Vinen calculated the driving force on a vortex for extreme type II superconductors. In this definition, we can arrive at the earlier force\textendash balance relation on a vortex in the TDGL scheme (with due correction of the driving force mentioned above). \\
{\it Model: } 
The TDGL equation
\begin{equation}
\gamma\left(\frac{\partial}{\partial t}+\frac{\iu e^*\Phi(\bm{r},t)}{\hbar}\right)\psi(\bm{r},t)=\xi^2 \left(\bm{\nabla}-\frac{\iu e^*\bm{A}(\bm{r},t)}{\hbar}\right)^2\psi(\bm{r},t)+\psi(\bm{r},t)-|\psi(\bm{r},t)|^2\psi(\bm{r},t) \label{eq: TDGL}
\end{equation}
has been used successfully to calculate flux flow Ohmic\cite{schmid,GK71} and Hall conductivities\cite{dorsey,kopnin93}. In Eq.~(\ref{eq: TDGL}), $\gamma$ denotes the complex relaxation time $\gamma=\gamma_1 +\iu \gamma_2$ and microscopic expressions have been given for $\gamma_1$ in Ref.~[\citen{GE68}] and $\gamma_2$ in Refs.~[\citen{fet}] and [\citen{ef}]. $\xi$ is the coherence length.  When the TDGL equation is valid, the Thomas\textendash Fermi screening length is much smaller than $\xi$ and thus the chemical potential can be regarded as spatially uniform and $\Phi$ can be regarded as the scalar potential. $\bm{A}$ denotes the vector potential. We take the macroscopic wavefunction $\psi$ to be dimensionless so that a spatially uniform solution for $\Phi=\bm{A}=0$ becomes $\psi=1$ up to an overall phase factor. 
An advantage of this model [Eq.~(\ref{eq: TDGL})] lies in its simplicity compared with the models for the dynamics for clean superconductors.  We should, however, keep in mind that the TDGL equation is justified microscopically only for dirty s-wave superconductors with a high concentration of paramagnetic impurities. 

For later convenience, we express the condensate wavefunction as $\psi=f(\bm{r},t){\rm e}^{\iu \chi(\bm{r},t)}$ in terms of the amplitude $f$ and the phase $\chi(\bm{r},t)$ and introduce the {\it gauge-invariant} scalar and vector potentials, respectively, as 
\begin{equation}
P=\Phi+\frac{\hbar}{e^*}\frac{\partial\chi}{\partial t},\quad
\bm{Q}=\bm{A}-\frac{\hbar\bm{\nabla}\chi}{e^*}.
\end{equation}
Here, $\chi$ and thus $P$ and $\bm{Q}$ are singular at the vortex center while $\Phi$, and $\bm{A}$ are regular. 
The electric field and magnetic field are, respectively, given by
$\bm{\varepsilon}=-\partial_t \bm{A}-\bm{\nabla}\Phi$ and $\bm{h}=\bm{\nabla}\times \bm{A}$,  
which are also expressed as
$\bm{\varepsilon}=-\partial_t \bm{Q}-\bm{\nabla}P$ and
$\bm{h}=\bm{\nabla}\times \bm{Q}$  
for any position but the vortex center. 
In terms of $f$, $P$, and $\bm{Q}$, Eq.~(\ref{eq: TDGL}) is rewritten as
\begin{equation}
\gamma\left(\frac{\partial}{\partial t}+\frac{\iu e^*P}{\hbar}\right)f=\xi^2 \left(\bm{\nabla}-\frac{\iu e^*\bm{Q}(\bm{r},t)}{\hbar}\right)^2f+f-f^3. \label{eq: TDGL-f-chi}
\end{equation}
We solve Eq.~(\ref{eq: TDGL-f-chi}) coupled with the Amp\`ere law 
\begin{equation}
\bm{\nabla}\times \bm{h}=\mu_0 \bm{j},\label{eq: Ampere}
\end{equation}
where $\mu_0$ is the permeability of vacuum. The current density $\bm{j}$ consists of two parts,  
\begin{equation}
\bm{j}=\bm{j}_{\rm s}+\bm{j}_{\rm n},\quad 
\bm{j}_{\rm s}=-\frac{f^2 \bm{Q}}{\mu_0\lambda^2},\quad 
\bm{j}_{\rm n}=\bm{\sigma}_{\rm n}\bm{\varepsilon}.\label{eq: jsjn}
\end{equation}
The subscripts ^^ ^^ n" and ^^ ^^ s" stand for normalfluid component and superfluid component, respectively. The symbol $\lambda$ denotes the penetration depth and \textcolor{black}{
\begin{equation}
\bm{\sigma}_{\rm n}=
\begin{pmatrix}
\sigma^{\rm O}_{\rm n}&\sigma^{\rm H}_{\rm n}& 0\\
-\sigma^{\rm H}_{\rm n}&\sigma^{\rm O}_{\rm n}& 0\\
0 & 0& \sigma^{\rm O}_{\rm n}\\
\end{pmatrix} 
\end{equation}}
is the conductivity tensor in the normal state. The superscripts ^^ ^^ O" and ^^ ^^ H" stand for Ohmic and Hall conductivities, respectively. 
We have dropped the displacement current term (Maxwell term) in Eq.~(\ref{eq: Ampere}) because we will consider the flux flow state in dc external current and the EM fields can be regarded as quasi-static\cite{LLem-continuummedia}. Equation~(\ref{eq: Ampere}) implies that $\nabla\cdot\bm{j}=0$
and the electric charge distribution is static. Note that we have already ignored the difference between the electrostatic chemical potential and the chemical potential and thus ignored the equilibrium charge inhomogeneity.  
Equations (\ref{eq: TDGL-f-chi})\textendash (\ref{eq: jsjn}) form a system of equations. 
For later convenience, we rewrite Eq.~(\ref{eq: TDGL-f-chi}) further. The real part of Eq.~(\ref{eq: TDGL-f-chi}) is given by
\begin{equation}
\gamma_1 \frac{\partial f}{\partial t} -\gamma_2\frac{e^* P f}{\hbar}=\xi^2\nabla^2 f-\left(\frac{e^*\xi \bm{Q}}{\hbar}\right)^2 f+f -f^3.\label{eq: TDGL-re}
\end{equation}
The imaginary part of Eq.~(\ref{eq: TDGL-f-chi}) yields
\begin{equation}
\frac{\partial \rho_{\rm s}}{\partial t}+\nabla\cdot\bm{j}_{\rm s}=\frac{\gamma_1 f^2 P}{\mu_0 \lambda^2\xi^2},\label{eq: TDGL-im}
\end{equation}
where we introduce the auxiliary notation
\begin{equation}
\rho_{\rm s}=-\frac{\gamma_2 \hbar f^2}{2\mu_0 e^*\lambda^2\xi^2}, 
\label{eq: rho-2}
\end{equation}
for later convenience. \textcolor{black}{Equation~(\ref{eq: TDGL-im})} leads us to interpret $\rho_{\rm s}$ as the superfluid component of the charge density and the right-hand side as the rate of conversion from normal to superfluid components. On the basis of this observation, we use the notation
\begin{equation}
\left(\frac{\rmd \rho_{\rm s}}{\rmd t}\right)_{\rm conv}\equiv\frac{\gamma_1 f^2 P}{\mu_0 \lambda^2\xi^2}
\label{eq: drhodt-2}
\end{equation}
on some occasions. \\
{\it Local balance relation of momentum density}: 
We find the local balance of the force density starting from the TDGL equation,
\begin{eqnarray}
&&\partial_t (-\rho_{\rm s}Q_\mu)+\partial_\nu \mathcal{P}_{\nu\mu}\nonumber\\
&&=\left(\rho_{\rm s}\bm{\varepsilon} +\bm{j}_{\rm s}\times \bm{h}\right)_\mu+\tilde{\gamma}_1\partial_\mu f \partial_t f+\left(-\frac{\rmd \rho_{\rm s}}{\rmd t}\right)_{\rm conv}Q_\mu,\label{eq: LRFB}
\end{eqnarray}
where we use the following notations: 
\begin{eqnarray}
&&{\cal P}_{\mu\nu}=-j^{\rm s}_{\nu}Q_{\mu}+\frac{2B_{\rm c}^2\xi^2 }{\mu_0}\partial_\nu f\partial_\mu f-\delta_{\mu\nu}{\cal F}_{\rm sn},\label{eq: Pmunu}\\
&&{\cal F}_{\rm sn}\equiv \frac{B_{\rm c}^2}{\mu_0}\left\{\xi^2|\bm{\nabla}f|^2+\left(\frac{\xi e^*\bm{Q} f}{\hbar}\right)^2-f^2+\frac{f^4}{2}\right\}+\rho_{\rm s}P,\label{eq: calFsn}\\
&&
\tilde{\gamma}_1=\frac{2B_{\rm c}^2}{\mu_0}\gamma_1,\\
&&
B_{\rm c}=\frac{|\phi_0^*|}{2\sqrt{2}\pi \lambda\xi}.
\end{eqnarray}
In Eq.~(\ref{eq: LRFB}), $-\rho_{\rm s}Q_\mu$ is regarded as the momentum density of a superfluid and ${\cal P}_{\mu\nu}$ the momentum flux tensor of a superfluid.
The first term on the right-hand side (RHS) of Eq.~(\ref{eq: LRFB}) represents the force on the superfluid component from the electromagnetic field and the second term represents the dissipation force due to the time variation of the modulus of the pair potential (^^ ^^ Tinkham mechanism"\cite{tinkham64}). The third force on the RHS represents the force density due to the conversion from normalfluid to superfluid components (or from superfluid to normalfluid components). 
In Eq.~(\ref{eq: Pmunu}), ${\cal P}_{\mu\nu}$ consists of the convection, quantum pressure, and Bernoulli terms. Expression (\ref{eq: calFsn}) yields the free energy difference between the superconducting state and the normal state per unit volume when the system is in equilibrium. 

Equation~(\ref{eq: LRFB}) follows from the TDGL equation without any additional assumption or approximations, as confirmed straightforwardly. We derive Eq.~(\ref{eq: LRFB}) in the supplementary online material for the benefit of interested readers\cite{som}. 
\ \\
{\it Force-balance relation for an isolated vortex in the flux flow state}: 
We derive the force-balance relation for the flux flow state of a single vortex from Eq.~(\ref{eq: LRFB}). We consider that physical quantities do not depend on $z$ (we will take the vortex axis parallel to the $z$-axis) and set $\gamma_2=0$ and $\sigma_{\rm n}^{\rm H}=0$ for simplicity. 

We rewrite the magnetic Lorentz force density $\bm{j}_{\rm s}\times \bm{h}$ as
\begin{eqnarray}
(\bm{j}_{\rm s}\times \bm{h})_\mu&=&(\bm{j}\times \bm{h})_\mu
-(\bm{j}_{\rm n}\times \bm{h})_\mu\nonumber\\
&=&\partial_\nu T_{\mu\nu}-(\bm{j}_{\rm n}\times \bm{h})_\mu\label{eq: js-Tmunu}
\end{eqnarray}
in terms of the Maxwell stress tensor $T_{\mu\nu}=\frac{1}{\mu_0}\left(h_\mu h_\nu -\frac12\delta_{\mu\nu}h^2\right)$. From Eqs.~(\ref{eq: LRFB}) and (\ref{eq: js-Tmunu}), we obtain 
\begin{equation}
\bm{F}_{\rm drv}+\bm{F}_{\rm env}=0\label{eq: FF}
\end{equation}
with 
\begin{eqnarray}
&&(\bm{F}_{\rm drv})_\mu \equiv\int_{\partial S}\left(-\mathcal{P}_{\mu\nu}+T_{\mu\nu}\right)n_\nu\rmd \ell\label{eq: Fdrv} \\
&&(\bm{F}_{\rm env})_\mu \equiv\int_{S}
\left(\tilde{\gamma}_1\left(\partial_\mu f \partial_t f-
(2\pi/\phi_0^*)^2 f^2 P Q_\mu\right)-\bm{j}_{\rm n}\times \bm{h}\right)_\mu\rmd S.\nonumber\\
\label{eq: Fenv}
\end{eqnarray}
Here, $\partial S$ is the boundary of a two-dimensional area $S$, which is perpendicular to the $z$-axis. $n_\mu$ denotes the $\mu=x,y$ component of the normal vector of $\partial S$. The direction of $\partial S$ is defined so that $\bm{e}_{z}\times \bm{n}$ has the same direction as the tangential vector of $\partial S$. $\rmd \ell$ is the line element along $\partial S$. \textcolor{black}{The driving force (\ref{eq: Fdrv}) acts as the sum of the hydrodynamic and magnetic pressures on the electrons in area $S$ through the surface $\partial S$.} 
In equilibrium, $\partial_t f=0$, $P=0$, $\bm{j}_{\rm n}=0$, and Eq.~(\ref{eq: FF}) reduces to $\bm{F}_{\rm drv}=0$ or 
\begin{equation}
\partial_\mu\left(-\mathcal{P}_{\mu\nu}+T_{\mu\nu}\right)=0\label{eq: local-balance}
\end{equation}
in the local form. Equation (\ref{eq: local-balance}) tells us that the internal stress is zero in equilibrium superconductors, as London pointed out\cite{london}.

We consider the flux flow state of an isolated vortex in the presence of a steady current in the $x$-direction and take $S$ as the disk with radius $R$ and the center $\bm{r}=0$ as the vortex axis. $\bm{F}_{\rm drv}$ and $\bm{F}_{\rm env}$  respectively turn out to be the driving and environment forces. 

$\bm{F}_{\rm env}$ is practically independent of $R$ when $R$ is much larger than $\xi$ because $\partial f/\partial t$ and $\partial_\mu f$ are localized around the region $r\lesssim \xi$ and $P$ is localized around the region $r\lesssim \zeta$, where the electric field penetration length $\zeta$ is given by the Hu\textendash Thompson length\cite{huthompson}$\zeta_{\rm HT}\equiv (\mu_0/\gamma_1)^\frac12 \lambda \xi$, which is of the order of $\xi$. For $R\gg \xi$, correspondingly, $\bm{F}_{\rm drv}$ is practically independent of $R$ and we will show later that 
\begin{equation}
\bm{F}_{\rm drv}\sim \bm{j}_{\rm tr}(0)\times \bm{\phi}^*_0. 
\label{eq: Fdrv-jtr} 
\end{equation}
Equation~(\ref{eq: FF}) together with Eqs.~(\ref{eq: Fdrv-jtr}) and (\ref{eq: Fenv}) becomes the well-known force\textendash balance relation\cite{kopnintext} if we identify $\bm{j}_{\rm tr}$ in the literature with $\bm{j}_{\rm tr}(0)$ in the present paper [compare with (12.16) on page 237 of Ref.~\citen{kopnintext}].  

\begin{figure}[t]
\begin{center}
\includegraphics[height = 5cm]{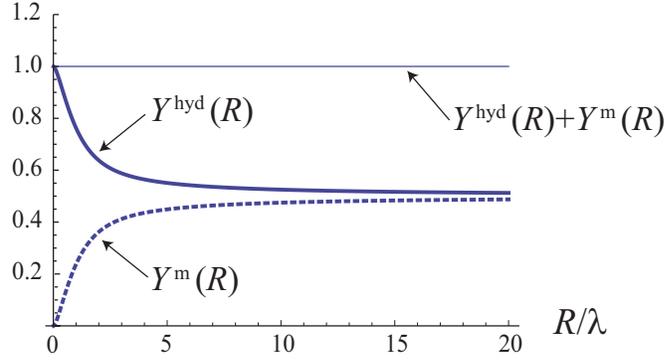}
\caption{(Color online) $R$-dependences of hydrodynamic driving force Eq.~(\ref{eq: Upsilon-hydro}) and magnetic force Eq.~(\ref{eq: Upsilon-em}). Here, we assume that $R\gg \xi$. }
\label{fig: Y}
\end{center}
\end{figure}

Equation~(\ref{eq: Fdrv-jtr}) is frequently referred to as the Lorentz force induced by an external current in the literature (e.g., see Ref.~\citen{kopnintext}).
We, however, find that the character of $\bm{F}_{\rm drv}$ depends crucially on the value of $R$. 
We define 
\begin{eqnarray}
-\int_{r=R}\mathcal{P}_{\mu\nu}n_\nu\rmd \ell&\equiv& 
(\bm{j}_{\rm tr}(0)\times \bm{\phi}_0^*)_\mu Y^{\rm hyd}(R)\label{eq: FMag} \\
\int_{r=R}T_{\mu\nu} n_\nu\rmd \ell&\equiv& 
(\bm{j}_{\rm tr}(0)\times \bm{\phi}_0^*)_\mu Y^{\rm m}(R)\label{eq: Fhydro} 
\end{eqnarray}
\textcolor{black}{so that we can see which part is dominant over the other. Explicit expressions for $Y^{\rm hyd}(R)$ and $Y^{\rm m}(R)$ are derived later. Figure~\ref{fig: Y} shows $Y^{\rm hyd}(R)$ and $Y^{\rm m}(R)$.}
 When $\xi\ll R \ll \lambda$, earlier works considered the force balance in the region with $R$ satisfying $\xi\ll R \ll \lambda$. In this case, $-\mathcal{P}_{\mu\nu}$ predominantly contributes to Eq.~(\ref{eq: Fdrv}) and thus it is legitimate to say that $\bm{F}_{\rm drv}$ is hydrodynamic (a part of the Magnus force) in nature.
When $R \gg \lambda$, in contrast, $-\mathcal{P}_{\mu\nu}$ and $\mathcal{T}_{\mu\nu}$ contribute to $\bm{F}_{\rm drv}$ equally. Thus, the nature of the driving force on the vortex depends on $R$. The crossover of the nature of the driving force with varying $R$ is a major result of this paper. \\
{\it Flux flow solutions of a single vortex}: 
As a prerequisite to derive expressions for $Y^{\rm hyd}(R)$ and $Y^{\rm m}(R)$, we summarize the flux flow solution of the TDGL equation.
For convenience, we express $\bm{\phi}^*_0=|\phi_0^*|{\rm sgn}(h_z(0))\hat{z}$ as
$\bm{\phi}^*_0=m\phi_0^* \hat{z}$,
where $m$ is the vorticity around $\bm{r}=0$;
\begin{equation}
m=\frac{1}{2\pi}\oint \rmd\bm{r}\cdot\bm{\nabla}{\rm Im}\ln\left(\psi/|\psi|\right).
\end{equation}
We denote by $X$ either $f$, $\bm{Q}$, $P$, $\bm{j}_{\rm s}$, or $\bm{j}_{\rm n}$. We seek the solutions to the TDGL and Amp\`ere equations in the form of 
\begin{equation}
X(\bm{r}, t)=X_0(\bm{r}-\bm{v}_{\rm L}t)+X_1(\bm{r})+\mathcal{O}(\bm{v}_{\rm L}^2), 
\end{equation}
where $X_0(\bm{r})$ is the solution for $\bm{v}_{\rm L}=0$, i.e., the static case. The second term on the RHS $X_1(\bm{r})$ is $\mathcal{O}(\bm{v}_{\rm L})$, in which the $t$-dependence is ignored because this dependence is $\mathcal{O}(\bm{v}_{\rm L}^2)$. 

In the equilibrium case, $\bm{v}_{\rm L}=0$, $P=0$, and $j_{\rm n0}=0$. For $r\gg \xi$, $f_0(r)\sim 1$  and the solution is known\cite{fetter} to be $\bm{j}_{s0}(\bm{r})=-\bm{Q}_0(\bm{r})/(\mu_0\lambda^2)$, $\bm{Q}_0(\bm{r})=Q_0(r)\bm{e}_\theta$, $\bm{h}_0(r)=h_0(r)\bm{e}_z$, $(\bm{e}_\theta=-\sin\theta\bm{e}_x +\cos\theta\bm{e}_y)$ with
\begin{eqnarray}
&&Q_0(r)\sim -\frac{m\phi_0^*}{2\pi\lambda}K_1\left(\frac{r}{\lambda}\right)\label{eq: Q0-asym-1}\\
&&h_0(r)\sim \frac{m\phi_0^*}{2\pi\lambda^2}K_0\left(\frac{r}{\lambda}\right).\label{eq: h0-asym-1}
\end{eqnarray}

In the region $\bm{r}$ satisfying $\xi\sim \zeta\ll r$, the first-order solutions have the asymptotic forms
\begin{subequations}
\begin{equation}
f_1\sim 0,\quad P_1\sim 0,\quad \bm{j}_{\rm n1}\sim 0\label{eq: jn1}
\end{equation}
and, accordingly, $\bm{j}_{\rm s1}$, $\bm{Q}_{\rm 1}$, and $h_{\rm 1}$  approach the bulk solution of the London and Amp\`ere equations in the absence of the vortex. 
\begin{eqnarray}
&&\bm{j}_{\rm s1}\sim \bm{j}_{\rm tr}(y)\\
&&\bm{Q}_{\rm 1}\sim -\mu_0\lambda^2\bm{j}_{\rm tr}(y)\\
&&\bm{h}_{\rm 1}\sim \mu_0 \lambda( j_+ {\rm e}^{y/\lambda} -j_- {\rm e}^{-y/\lambda})\bm{e}_z.\label{eq: h1}
\end{eqnarray}
\label{eq: 1st-order-bc}
\end{subequations}
{\it Magnetic part of the driving force $Y^{\rm m}(R)$}: 
We obtain with the accuracy of $\mathcal{O}(\bm{v}_{\rm L}^2)$
\begin{eqnarray}
&&\int_{r=R}T_{\mu\nu}n_{\nu}\rmd \ell\\
&=&-\frac{1}{\mu_0}\int_{r=R} \bm{h}_0\cdot\bm{h}_1 n_{\mu}\rmd \ell\\
&=&-h_0(R)R \lambda \int_{0}^{2\pi} 
\left(j_+ {\rm e}^{R\sin\theta/\lambda}-j_-{\rm e}^{-R\sin\theta/\lambda}\right)\left(\bm{e}_x \cos\theta +\bm{e}_y \sin\theta\right)_\mu\rmd \theta\nonumber\\
&=&-(j_+ +j_-)m\phi_0^* \frac{R}{\lambda}K_0\left(\frac{R}{\lambda}\right)
I_1\left(\frac{R}{\lambda}\right)\left(\bm{e}_y\right)_\mu.\label{eq: 12}
\end{eqnarray}
Here, $I_j(x)$ [$K_j(x)$] denotes the $j$th-order modified Bessel function of the first [second] kind. We have used the relation
\begin{equation}
\int_{0}^{\pi} {\rm e}^{x\cos\theta}\rmd \theta=\pi I_0(x),\quad I'_0(x)=I_1(x).\label{eq: ModifiedBessel} 
\end{equation}
With the use of $-(j_+ +j_-)m\phi_0^* \bm{e}_y=\bm{j}_{\rm tr}(0)\times \bm{\phi}_0^*$, we obtain
\begin{equation}
Y^{\rm m}(R)\simeq\frac{R}{\lambda}K_0\left(\frac{R}{\lambda}\right)
I_1\left(\frac{R}{\lambda}\right)\label{eq: Upsilon-em}.
\end{equation}
{\it Hydrodynamic part of the driving force $Y^{\rm hyd}(R)$}: Among the terms in~(\ref{eq: Pmunu}), the contribution proportional to $\mathcal{O}(\bm{v}_{\rm L})$ should be retained. We first consider the convection term $-j_{{\rm s},\nu}Q_{\mu}$ and replace it by $-j_{{\rm s0},\nu}Q_{1\mu}-j_{{\rm s1},\nu}Q_{0\mu}$. The term $-j_{{\rm s0},\nu}Q_{1\mu}$ does not contribute to  $Y^{\rm hyd}(R)$  because $\bm{j}_{\rm s0}$ is tangential to the circumference and perpendicular to the normal vector $\bm{n}$. The term $-j_{{\rm s1},\nu}Q_{0\mu}$ yields 
\begin{eqnarray}
&&-\int_{r=R}n_{\nu}j_{{\rm s1},\nu}Q_{0\mu}\rmd \ell\nonumber\\
&&=
-(j_+ +j_-)R Q_0(R)\left(\bm{e}_y\right)_\mu 
\int_{0}^{2\pi} \cos^2\theta\cosh\left(\frac{R}{\lambda}\sin\theta\right)\rmd\theta
\label{eq: Pmunu-1}
\end{eqnarray}
and contributes to $Y^{\rm hyd}(R)$. 
The second term on the RHS of Eq.~(\ref{eq: Pmunu}) (quantum pressure)
is negligible. We then consider the third term (the ^^ ^^ Bernoulli term"): 
$\mathcal{O}(\bm{v}_{\rm v})$ terms in $\mathcal{F}_{\rm sn}$ are given by
\begin{eqnarray}
&&\frac{2B_{\rm c}^2}{\mu_0}\xi^2 \bm{\nabla}\cdot\left(f_1 \bm{\nabla} f_0\right)-\bm{Q}_0 \cdot\bm{j}_{{\rm s},1}\label{eq: Fsn-1st}
\end{eqnarray}
with the use of the relation $\left(-\xi^2 \bm{\nabla}^2 f_0 +\left(\frac{\xi e^*\bm{Q}_0}{\hbar}\right)^2 f_0 -f_0 +f_0^3\right)=0$.
The first term in Eq.~(\ref{eq: Fsn-1st}) is negligible and thus
\begin{eqnarray}
&&\int_{r=R}\mathcal{F}_{\rm sn}n_\mu \rmd \ell\nonumber\\
&\sim&-\int_{r=R}\bm{Q}_0 \cdot\bm{j}_{{\rm s}1}n_\mu \rmd \ell\nonumber\\
&=&(j_+ +j_-)R Q_0(R)\left(\bm{e}_y\right)_\mu 
\int_{0}^{2\pi} \sin^2\theta\cosh\left(\frac{R}{\lambda}\sin\theta\right)\rmd\theta.\nonumber\\
\label{eq: Pmunu-2}
\end{eqnarray}
From Eqs.~(\ref{eq: ModifiedBessel}), (\ref{eq: Pmunu-1}), and (\ref{eq: Pmunu-2}), 
we obtain 
\begin{equation}
Y^{\rm hyd}(R)\simeq\frac{R}{\lambda}K_1\left(\frac{R}{\lambda}\right)
I_0\left(\frac{R}{\lambda}\right).\label{eq: Upsilon-hydro}
\end{equation}
{\it Nature of the driving force}: 
From Eqs.~(\ref{eq: Upsilon-em}) and (\ref{eq: Upsilon-hydro}) and the relation
\begin{equation}
xK_1(x)I_0(x)+xK_0(x)I_1(x)=1,
\end{equation}
we arrive at $Y^{\rm hyd}(R)+Y^{\rm m}(R)\simeq 1$, i.e., 
Eq.~(\ref{eq: Fdrv-jtr}). This fact justifies the definition [Eq.~(\ref{eq: Fdrv})] of the driving force as the surface integral of the sum of the momentum flux tensor and the Maxwell stress tensor. \\
{\it Discussion}: 
The results in the present paper rely on the following two assumptions:\\
{[\bf a1]} Far from the vortex center, a dissipation-free region exists where the stress-free condition (\ref{eq: local-balance}) holds. \\
{[\bf a2]} The external magnetic field generated by the current or magnets outside superconductors is negligible so that we are allowed to consider only the magnetic field generated by $\bm{j}_{\rm s}$ and $\bm{j}_{\rm n}$. \\
When the two assumptions are justified, our argument is valid.
We expect that our conclusion also applies to an isolated vortex near $H_{\rm c1}$ for both dirty and clean superconductors. 
We also consider that our conclusion is applicable to the nature of the driving force on an isolated {\it pinned} vortex, which has been considered in Refs.~\citen{friedel,deGennes,dxchen98,Narayan}, where [{\bf a1}] and [{\bf a2}] both held.  
In contrast, the two assumptions [{\bf a1}] and [{\bf a2}] are not necessarily justified for vortex lattices in the flux flow state and pinned state; in particular, near $H_{\rm c2}$, neither of the two holds. A future problem is to investigate the force on vortices in vortex lattices in the flux flow state and pinned state.\\
\textcolor{black}{{\it Conclusion}: We studied the nature of the driving force on an isolated vortex while taking account of the spatial variation of the transport current. The driving force on the vortex consists of hydrodynamic and magnetic pressures. The sum of the two pressures is the only physically meaningful force in the sense that hydrodynamic and magnetic pressures depend on the choice of the area of superconducting electrons that surround the vortex but their sum does not.} 
\ \\
{\it Acknowledgments}: This work was supported by JSPS KAKENHI Grant Number 15K05160. The authors thank Y.~Masaki and S.~Hoshino for their critical reading of the manuscript.

\end{document}